\documentclass[twocolumn,showpacs,preprintnumbers,amsmath,amssymb,prb,superscriptaddress]{revtex4-2}
\usepackage[dvipdfmx]{graphicx}% Include figure files
\usepackage{dcolumn}% Align table columns on decimal point
\usepackage{bm}% bold math
\usepackage{color} % \color{red} \color{black}
\usepackage{ulem} % \sout - single out

\usepackage{xspace}

\newcommand{\Ang}{\AA\xspace}

\newcommand{\Smtri}{Sm$^{3+}$\xspace}
\newcommand{\Smdi}{Sm$^{2+}$\xspace}
\newcommand{\dR}{$\Delta R/R$\xspace}

\newcommand{\hn}{$h\nu$\xspace}

\newcommand{\mJ}{mJ/cm$^{2}$\xspace}

\begin{document}

\title{
Photo-induced phase transition on black samarium monosulfide
}
\author{Hiroshi~Watanabe}
\affiliation{Graduate School of Frontier Biosciences, Osaka University, Suita 565-0871, Japan}
\affiliation{Department of Physics, Graduate School of Science, Osaka University, Toyonaka 560-0043, Japan}
\author{Yusuke~Takeno}
\affiliation{Department of Physics, Graduate School of Science, Osaka University, Toyonaka 560-0043, Japan}
\author{Yusuke~Negoro}
\affiliation{Department of Physics, Graduate School of Science, Osaka University, Toyonaka 560-0043, Japan}
\author{Ryohei~Ikeda}
\thanks{Present address: Department of Advanced Materials Science, Graduate School of Frontier Sciences, The University of Tokyo, Kashiwa 277-8561, Japan}
\affiliation{Department of Physics, Graduate School of Science, Osaka University, Toyonaka 560-0043, Japan}
\author{Yuria~Shibata}
\affiliation{Department of Physics, Graduate School of Science, Osaka University, Toyonaka 560-0043, Japan}
\author{Yitong~Chen}
\affiliation{Department of Physics, Graduate School of Science, Osaka University, Toyonaka 560-0043, Japan}
\author{Takuto~Nakamura}
\affiliation{Graduate School of Frontier Biosciences, Osaka University, Suita 565-0871, Japan}
\affiliation{Department of Physics, Graduate School of Science, Osaka University, Toyonaka 560-0043, Japan}
\author{Kohei~Yamagami}
\thanks{Present address: Japan Synchrotron Radiation Research Institute, Sayo 679-5198, Japan}
\affiliation{Institute for Solid State Physics, The University of Tokyo, Kashiwa 277-8581, Japan}
\author{Yasuyuki~Hirata}
\affiliation{Department of Applied Physics, National Defense Academy, Yokosuka 239-8686, Japan}
\author{Yujun~Zhang}
\affiliation{Department of Material Science, Graduate School of Science, University of Hyogo, Ako 678-1297, Japan}
\author{Ryunosuke~Takahashi}
\affiliation{Department of Material Science, Graduate School of Science, University of Hyogo, Ako 678-1297, Japan}
\author{Hiroki~Wadati}
\affiliation{Department of Material Science, Graduate School of Science, University of Hyogo, Ako 678-1297, Japan}
\author{Kenji~Tamasaku}
\affiliation{RIKEN SPring-8 Center, Sayo 679-5148, Japan}
\author{Keiichiro~Imura}
\affiliation{Institute of Liberal Arts and Sciences, Nagoya University, Nagoya 464-8601, Japan}
\author{Hiroyuki~S.~Suzuki}
\affiliation{Institute for Solid State Physics, The University of Tokyo, Kashiwa 277-8581, Japan}
\author{Noriaki~K.~Sato}
\affiliation{Center for General Education, Aichi Institute of Technology, Toyota 470-0392, Japan}
\author{Shin-ichi~Kimura}
\email{sk@kimura-lab.com}
\affiliation{Graduate School of Frontier Biosciences, Osaka University, Suita 565-0871, Japan}
\affiliation{Department of Physics, Graduate School of Science, Osaka University, Toyonaka 560-0043, Japan}
\affiliation{Institute for Molecular Science, Okazaki 444-8585, Japan}
\date{\today}
\begin{abstract}
To investigate the role of the excitons for the origin of the pressure-induced phase transition (BGT) from the black-colored insulator (BI) to the golden-yellow-colored metal (GM) of samarium monosulfide (SmS), optical reflectivity, Sm~$3d$ X-ray absorption spectroscopy (XAS), and X-ray diffraction (XRD) with the creation of excitons by photoexcitation (PE) are reported.
In the pump-probe reflectivity measurement, following a huge reflectivity change of about 22~\%, three different relaxation times with a vibration component were observed.
The fast component with the relaxation time ($\tau$) of less than 1~ps is due to the excitation and relaxation of electrons into the conduction band, and the slowest one with $\tau > {\rm several}~100$~ps originates from the appearance of the photo-induced (PI) state.
The components with $\tau \sim 10$~ps and vibration originate from the appearance of the PI state and the interference between the reflection lights at the sample surface and the boundary between the BI and PI states, suggesting that the electronic structure of the PI phase is different from that of the BI state. 
%It appears with the middle relaxation component with $\tau \sim 10$~ps, a transient region to the PI state.
XAS spectra indicate that the Sm mean valence is shifted from the \Smdi dominant to the intermediate between \Smdi and \Smtri by PE but did not change to that of the GM phase across BGT, consistent with the reflectivity data.
The XRD result after PE shows that the PI state has much less lattice contraction than the GM phase.
These results suggest that the BGT cannot be achieved solely by creating excitons after PE but requires other effects, such as a lattice contraction.
\end{abstract}

%
%\pacs{71.27.+a, 78.20.-e}% PACS, the Physics and Astronomy
%%%%%%%%%%%%%%%%%%%%%%%%%%%%%%%%%%%%%%%%%%%%%%%%%%%%%%%%%%%%
\maketitle
%
%%%%%%%%%%%%%%%%%%%%%%%%%%%%%%%%%%%%%%%%%%%%%%%%%%%%%%%%%%%%
\section{Introduction}

Materials with strong electron correlation, so-called strongly correlated electron systems, exhibit characteristic physical properties caused by intense electron-electron interactions, such as magnetic ordering, heavy-fermion states, and non-BCS superconductivity.
Since these states are based on a balance among interactions of electrons, a phase transition occurs when the balance is disrupted.
The phase transitions can be controlled with external perturbations.
For example, the transition between a localized magnetically ordered state and an itinerant heavy fermion state in heavy fermion systems can be induced by external pressure, a magnetic field, or chemical pressure due to substituting substances with different ionic radii~\cite{Si2010-ik}.
One method of disrupting the balance is the excitation by intense pulsed light~\cite{Nasu2004-px}.

It is known that electrons are excited by pulsed light and appear in a metastable state, different from a ground state.
For example, a Josephson plasma state is observed at room temperature by photoexcitation (PE) in cuprate superconductors, suggesting a realization of room temperature superconductivity~\cite{Cavalleri2018-su}.
In (EDO-TTF)$_2$PF$_6$, for another example, the charge-ordered state, which appears only by PE, originates from a giant ultrafast photo-induced (PI) change in physical properties~\cite{Chollet2005-vm}.
Such PI phase transitions appear when the quantum states of materials are located near phase boundaries.
One such candidate is samarium monosulfide (SmS).

For more than 50 years, it has been known that the state of SmS changes from black-colored insulator (BI) to golden-yellow-colored metal (GM) by the first-order black-to-golden transition (BGT) under pressure~\cite{Jayaraman1970-ky,Maple1971-av,Kirk1972-ra,Bader1973-gj,Batlogg1974-gg,Matsubayashi2007-bq,Mizuno2008-lg,Matsubayashi2007-ej,Imura2015-mq,Imura2020-zq}.
However, the origin of the phase transition is not yet elucidated.
One theory that has been proposed is that of excitonic instability~\cite{Cloizeaux1965-gu,Varma1976-op}.
In this theory, when the exciton level in the energy gap is reduced by pressure to a lower energy than the energy gap, electrons in the valence band automatically flow into the exciton levels and become metallic.
Numerical calculations using the periodic Anderson model suggest that exciton density increases near the BGT boundary, resulting in excitonic instability~\cite{Watanabe2021-rx}.
It is possible that PE can produce a massive number of excitons and become a trigger of BGT.
PE also possibly induces metastable states, which are not reached by an external pressure.

Time-resolved photoemission spectroscopy has been applied to the change in the valence band of SmS due to PE~\cite{Chen2022-fn}.
It was found that a surface photovoltage effect occurred immediately after PE, and then the indirect energy gap, which is usually about 100~meV, decreased to about 50~meV and saturated.
Although the minimum size of the energy gap is equal to the indirect gap energy just before pressure-induced BGT, the photoemission spectrum does not change significantly from that of BI.
This fact suggests that the overall electronic state after PE is similar to that of BI without PE, but only the energy gap size shrinks.
It needs to be clarified how the electronic state change observed in the time-resolved photoemission spectroscopy is related to the lattice and valence changes.

PI phase transition of SmS has been reported in other literatures~\cite{Pohl1974-qf,Kitagawa2003-ez}.
Crystal strain induced by the substrate for thin films and the relaxation and/or damage by laser heating might play an essential role in the phase transitions.
However, the detailed mechanism related to electronic excitation and its relation to BGT still needs to be clarified.
Here, we investigated the changes in the physical properties of SmS when irradiated with pulsed light without damaging the sample.
Specifically, we investigated changes in macroscopic optical constants and their expansion in the crystal, as well as changes in valence and lattice constants, by measuring an optical reflectivity, X-ray absorption spectroscopy (XAS) at the Sm~$3d$ core level, and X-ray diffraction (XRD) after PE.
These results suggest that PE induces changes in the electronic structure and optical constants, but the lattice constant does not change from the BI phase.
The results suggest that additional effects, such as lattice contraction, are needed for the BGT.

%%%%%%%%%%%%%%%%%%%%%%%%%%%%%%%%%%%%%%%%%%%%%%%%%%
\section{Experimental}

A single crystal was grown using a vertical Bridgman method in a high-frequency induction furnace as described in a literature~\cite{Matsubayashi2007-ej}.
Measured samples along the (001) plane were obtained by a cleaving method.

Pump-probe reflectivity was measured using a Ti:Sapphire pulse laser with a regenerative amplifier (photon energy ($h\nu$): $\sim 1.55$~eV, wavelength ($\lambda$): $\sim 800$~nm, maximum average power: $\sim 1$~mJ, reputation: $1$~kHz, pulse width: $\sim 50$~fs) at the temperature of 300~K.
Probe light was detected with a Si photodiode.
An optical chopper at 500~Hz modulates the excitation light to obtain high accuracy of the intensity change between with and without the excitation light.
The diffraction intensities are accumulated using a lock-in amplifier.

The photoexcited XAS and XRD experiments were performed at BL07LSU~\cite{Yamamoto2014-aa} and BL19LXU~\cite{Tanaka2000-uw,Oura2018-aa}, respectively, of SPring-8 using a several-bunch mode, namely the H mode (11/29-filling + 1-bunch).
The pump-probe systems for the XRD and XAS are almost the same as the reflectivity system, but the probe light is an X-ray (pulse width $\sim 50$~ps) from SPring-8, where the laser pulses are synchronized to single-bunch X-rays.
The XAS spectra were accumulated at around the Sm~$3d$ absorption edge at \hn$\sim 1080$~eV using the time-resolved XAS setup~\cite{Takubo2017-kz,Yokoyama2019-nt} at $58$~K.
The XRD measurement used the 12.4-keV photons and an X-ray camera SOPHIAS~\cite{Hatsui2015-rl}, and was performed only at $300$~K.
The Bragg (400) reflection peak was selected owing to the limitation of the experimental setup.
In all experiments, the spot size of the pump light on samples was 1--2~mm in diameter.

%%%%%%%%%%%%%%%%%%%%%%%%%%%%%%%%%%%%%%%%%%%%%%%%%%
%%%%%%%%%%%%%%%%%%%%%%%%%%%%%%%%%%%%%%%%%%%%%%%%%
\section{Results}

%%%%%%%%%%%%%%%%%%%%%%%%%%%%%%%%%%%%%%%%%%%%%%%%%%
%%%%%%%%%%%%%%%%%%%%%%%%%%%%%%%%%%%%%%%%%%%%%%%%%
\subsection{Time-resolved reflectivity}

When BGT causes the transition from an insulating electronic state to a metallic one, or the energy gap narrows in an insulating electronic structure, a significant reflectivity change below the plasma edge, or the energy gap, is expected.
This means that when BGT or the energy gap change is induced by PE, a significant change in optical constants, including reflectivity, and subsequent relaxation will appear.
Therefore, we measured the time variation of reflectivity at \hn$\sim 1.55$~eV ($\lambda \sim 800$~nm), which is lower energy than the plasma-edge energy of about 3~eV in GM~\cite{Travaglini1984-pq}, by excitation with pulsed laser of the same energy.
Since the excitation energy is higher than that of the direct energy gap of BI~\cite{Kimura2008-ty}, electrons in the valence bands are expected to be excited into the conduction bands.

%%%%%%%%%%%%%% FIG. R1. tr Reflectivity %%%%%%%%%%%%%%%%%%%%
\begin{figure}[t]
\begin{center}
\includegraphics[width=0.48\textwidth]{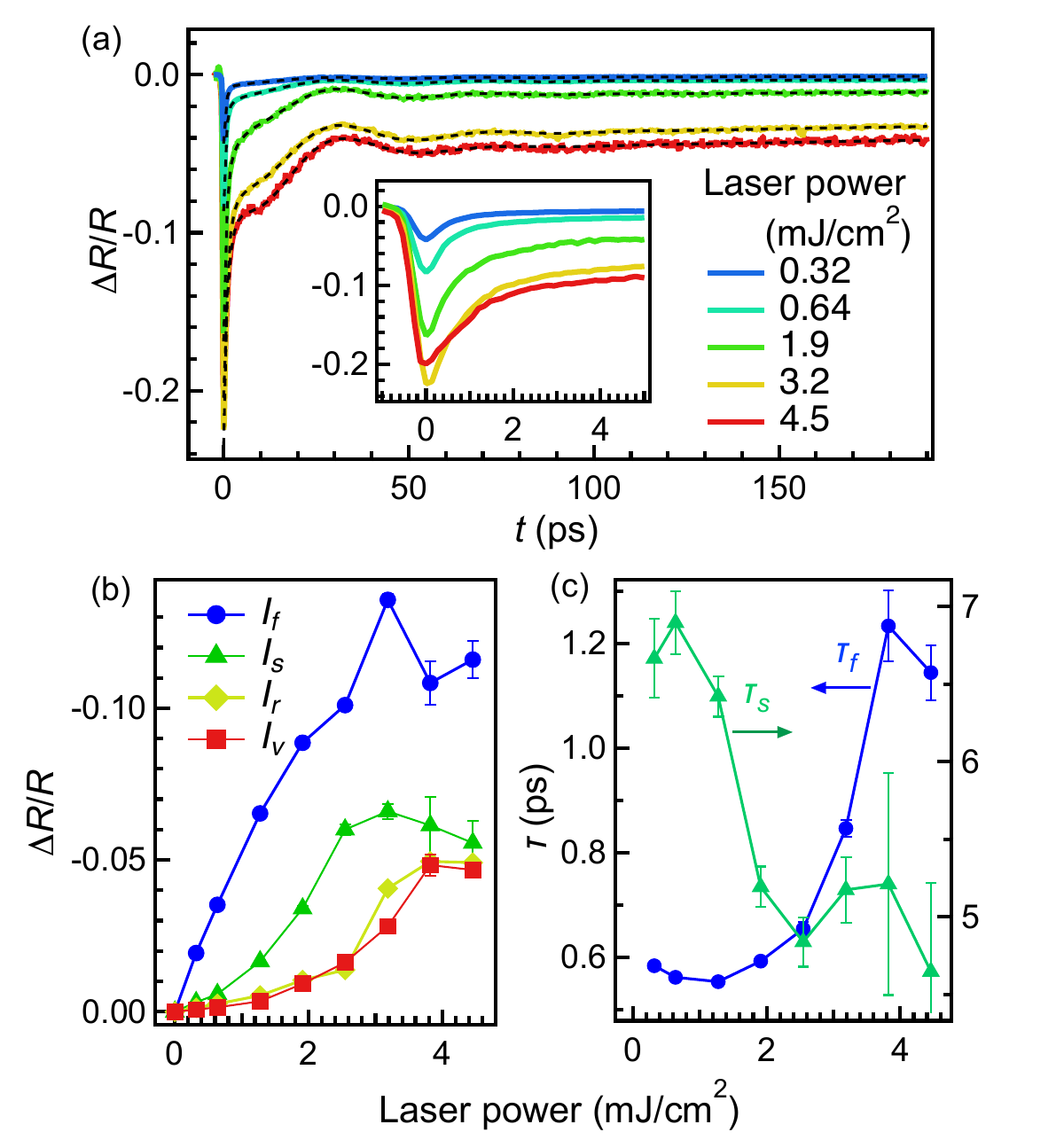}
\end{center}
\caption{
Time-dependent reflectivity of BI at 800-nm wavelength ($h\nu \sim 1.55$~eV) after PE at 300~K and the fitting results using Eq.~\ref{eq:decay}. 
(a)~Time and excitation-power dependences of the reflectivity change ($\Delta R/R$).
The fitting curves are shown with black dashed lines.
(Inset)~The enlargement of the laser-power dependence of the time structure of the fast component.
(b)~Excitation-laser-power dependence of each component's intensity obtained by the fitting.
The components' symbols can be referred to in the text.
(c)~Excitation-laser-power dependence of relaxation time of the fast ($\tau_f$) and slow ($\tau_s$) components obtained by the fitting.
Error bars in (c) and (d) are the standard deviations of the fitting parameters evaluated using a least square method.
}
\label{fig:R1}
\end{figure}
%%%%%%%%%%%%%%%%%%%%%%%%%%%%%%%%%%%%%%%
Figure~\ref{fig:R1}(a) shows the time and excitation-power dependence of reflectivity change ($\Delta R/R$) at \hn~=~1.55~eV.
The horizontal axis is the elapsed time since the pump pulse was irradiated at $t = 0$.
The maximum magnitude of \dR was about 22~\% just after PE with the 3.2-\mJ laser power, and a substantial decrease in the magnitude of \dR was observed. 
After that, the magnitude of \dR decreased by a fast relaxation of less than 1~ps and a slow relaxation of $\sim 10$~ps.
Finally, it became a very slow relaxation (here, described as a ``residual component'') that could not fully relax within the measurement time range of 190~ps.
This implies that the excited electrons do not directly relax to the initial state, but a metastable PI state appears.
The PI state returns to the initial state before the next PE since the magnitude of \dR returns to zero at 2~ms after PE, the time width of 500~Hz of the repetition frequency used in the reflectivity measurement.
Furthermore, vibrations of the \dR were observed; the vibration amplitude gradually decreases and becomes invisible at $\sim 100$~ps after PE.

When the irradiated laser power was less than 3.8~\mJ, the samples were not damaged, and the repeated experiment showed the same results.
However, when PE with a higher intensity than 4.0~\mJ, \dR was not reproduced due to damage to the samples.
As shown in the inset of Fig.~\ref{fig:R1}(a), with increasing the pump laser power, the magnitude of \dR increases in the laser power below 3.2~\mJ and is almost saturated at 4.5~\mJ.
The saturation is considered to originate from the sample damage.

\begin{figure}[t]
\begin{center}
\includegraphics[width=0.48\textwidth]{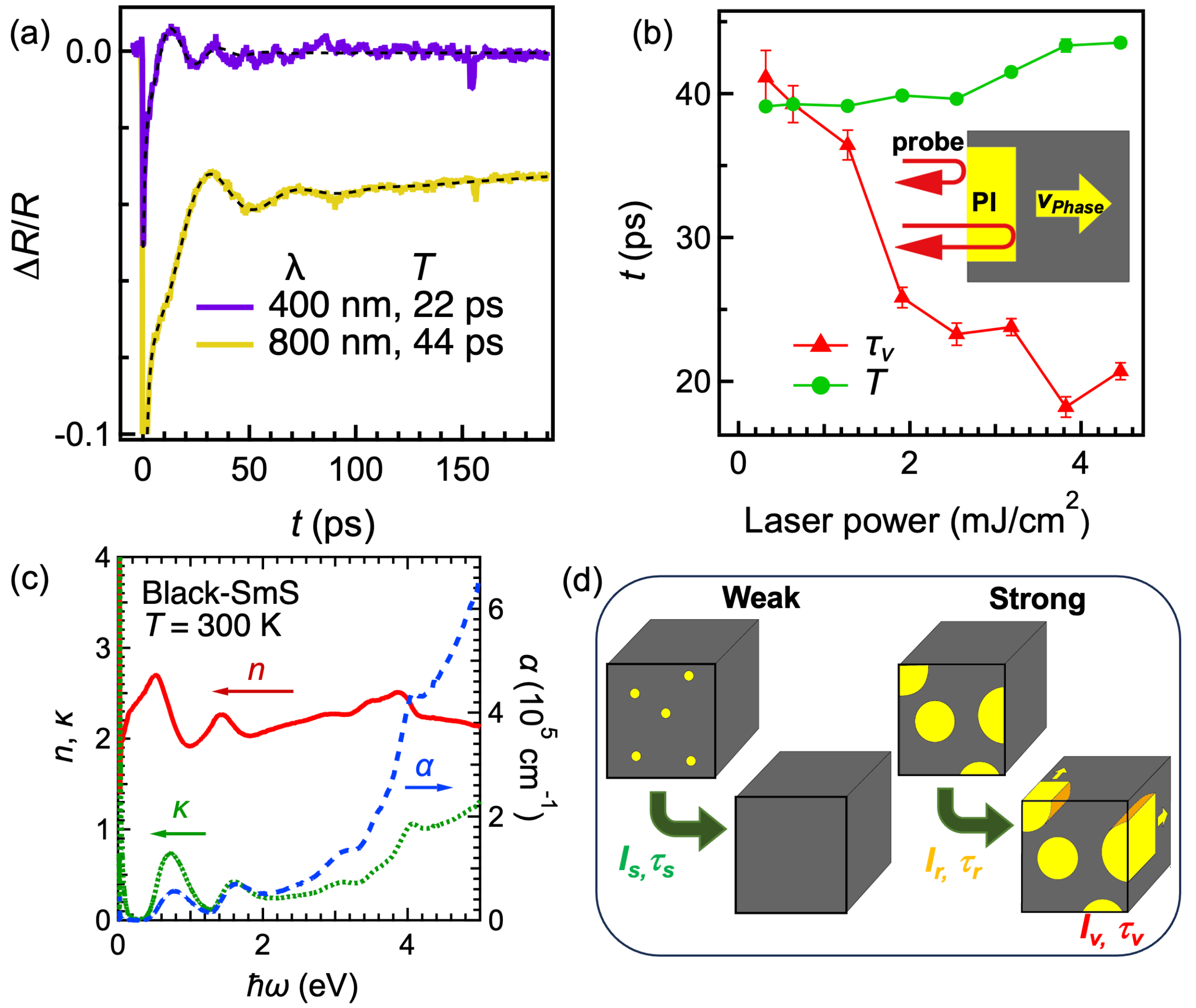}
\end{center}
\caption{
(a)~Time-dependent reflectivity change (\dR) at 400-nm wavelength compared to 800-nm.
(b)~Excitation-laser-power dependence of relaxation time $\tau_v$ and the period $T$ of the vibration component.
Error bars are the standard deviations of the fitting parameters evaluated using a least square method.
The inset illustrates a schematic figure of the origin of the interference between reflections from the sample surface and the boundary between the PI and BI phases.
The PI phase expands to the sample inside with a velocity $v_{phase}$.
(c)~Refractive index $n$, extinction coefficient $\kappa$, and absorption coefficient $\alpha$ spectra of BI derived from the Kramers-Kronig analysis of the reflectivity spectrum~\cite{Mizuno2008-lg}.
(d)~Schematic figure of the creation and annihilation of the PI phase at the weak and strong excitation intensities.
In the weak excitation case, small isolated PI domains are created but disappear soon, but in the strong excitation case, large cooperative PI domains appear and expand inside the sample.
}
\label{fig:R2}
\end{figure}

We performed a component deconvolution using the following assumptions to investigate the origin of the time structures in \dR.
The \dR is considered to consist of four components; 
a fast component (the intensity: $I_f$, the relaxation time: $\tau_f$) that relaxes in about 1~ps, 
a slow component ($I_s$, $\tau_s$) in about 10~ps, 
a vibration component ($I_v$, $\tau_v$, the period: $T$, the phase shift: $\phi$) that gradually decays during relaxation, 
and 
a residual component ($I_r$, $\tau_r$) that remains a change in reflectivity after a longer time than the accessible time delay of 190~ps.
Therefore, we fitted \dR curves with the following function containing the assumed three exponential relaxation components and one damping vibration component:
\begin{eqnarray}
&-\dfrac{\Delta R}{R} = I_{f}\exp \left[-\dfrac{t}{\tau_{f}}\right]+I_{s}\exp \left[-\dfrac{t}{\tau_{s}}\right] \nonumber \\
&+I_{v}\exp \left[-\dfrac{t}{\tau_{v}}\right]\sin\left[\dfrac{2\pi t}{T}+\phi\right]+I_{r}\exp \left[-\dfrac{t}{\tau_{r}}\right]
\label{eq:decay}
\end{eqnarray}
The fitting results of the above equation are shown by black dashed lines in Fig.~\ref{fig:R1}(a), and the critical obtained parameters are shown in Figs.~\ref{fig:R1}(b, c) and \ref{fig:R2}(b).

The PE-power-dependent intensities of all components are shown in Fig.~\ref{fig:R1}(b).
The fast component $I_f$ shows different behavior from other components, linearly increasing initially and decreasing above 3.2~\mJ.
In other components, however, intensities nonlinearly increase below 2.6~\mJ for the slow component $I_s$ and 3.8~\mJ for the vibration component $I_v$ and residual component $I_r$.
There seems to be a threshold at about 2.2~\mJ in $I_v$ and $I_r$, indicating that the two components' behavior is very similar.
This fact suggests that the vibration component is related to the residual component, the PI state.
$I_s$ rises earlier than the other two and saturates above 2.6~\mJ, lower than other saturation power.
All the data above 4.0~\mJ differ from the values below because of the damage by the PE.

Firstly, we discuss the origin of the fast component.
The intensity $I_f$ linearly increases with increasing excitation intensity but suddenly decreases above 2.6~\mJ.
The linear increase indicates that the number of excited electrons is proportional to the number of applied photons at low PE intensities and begins to saturate at around 2.6~\mJ.
The \dR decreases significantly, and this relaxation $\tau_f$ occurs on a time scale of about 1~ps, as shown in Fig.~\ref{fig:R1}(c).
The fact suggests that the fast component is due to the suppression of optical absorption as the photo-excited electrons are filled at the excited state.
In addition, $\tau_f$ shows a very strong nonlinearity to the excitation laser power and rapidly increases above 2.6~\mJ, where $I_f$ begins to saturate.
This phenomenon is presumably due to the limited relaxation pass, which causes many electrons in the excited state to take longer to decay to the initial state.

Secondly, the origin of the slow component is discussed.
The intensity $I_s$ quadratically increases and then saturates at $\sim 2.6$~\mJ.
This relaxation occurs on the order of 10~ps (shown in Fig.~\ref{fig:R1}(c)), unlike the fast component.
It is thought to involve the formation of the PI phase.
The laser power induces a fraction of the PI phase on the sample surface, but the PI state disappears soon when the laser power is not as high as the saturation intensity of $\sim$~2.6~\mJ.
At a higher intensity than 2.6~\mJ, the PI state eventually shifts to the residual component because $I_r$ rapidly increases above the laser power.
The vibration component originates from the expanding PI phase inside the sample.
The reason is discussed below.

%%%%%%%%%%%%%% FIG. R2. tr Reflectivity %%%%%%%%%%%%%%%%%%%%
%%%%%%%%%%%%%%%%%%%%%%%%%%%%%%%%%%%%%%%

Finally, we discuss the origin of the vibration component of this sample.
As seen in Fig.~\ref{fig:R2}(a), the vibration period has no excitation-power dependence, {\it i.e.}. the vibration period is a fixed value originating from the material property.
Such vibration is usually assigned as a coherent phonon (CP)~\cite{Dekorsy1999-xh}.
To clarify whether the vibration originates from CP, the wavelength of the probe laser was changed to 400~nm by using a second-harmonic generation with a BBO crystal.
In the case of CP, the vibration period originating from the eigenvalues of phonons does not change when the color is changed.
However, as seen in Fig.~\ref{fig:R2}(a), the period using the 400-nm wavelength probe is halved from the 800-nm wavelength probe.
Because BI has almost the same refractive index of the 400-nm wavelength as that of the 800-nm wavelength shown in Fig.~\ref{fig:R2}(c), the period is strongly related to the interference of the probe light.
This fact means that the origin is not CP.
The interference originates from the reflections at the sample surface and the boundary between the PI and BI phases propagating into the sample inside, as illustrated in the inset of Fig.~\ref{fig:R2}(b).
The reflection at the PI--BI boundary originates from the different refractive index of the PI state from that of BI.
This fact suggests that the optical constants of the PI state are not the same as those of BI, {\it i.e.}, the electronic structure of the PI state is changed by PE.

From the fitting, the phase shift $\phi$ was determined as almost $0$.
This suggests that the relative refractive index at the interface between the PI and BI state ($n_{PI}/n_{BI}$) is larger than $1$, which is the opposite relation between the vacuum and the PI state ($1/n_{PI} < 1$).
Since $n_{BI}$ is about $2$ from Fig.~\ref{fig:R2}(c), $n_{PI}$ must be $1 < n_{BI} (\sim 2) < n_{PI}$.

The propagation velocity of the PI phase, $v_{phase}=\lambda/(2n T)$, can be evaluated as about 5~km/s from the parameters of $T \sim 40$~ps in the 800-nm probe, and $n$ of BI at the 800-nm wavelength is about 2 as shown in Fig.~\ref{fig:R2}(c).
This $v_{phase}$ corresponds to the speed of sound in a solid, indicating that the PI phase expands into the sample inside by the sound velocity. 

On the other hand, the light penetration length ($\ell$) is estimated as about 200~nm from the relation of $\ell = 2 \tau_v \cdot v_{phase}/n$, where $\tau_v \sim 40$~ps, $v_{phase} \sim 5$~km/s and $n \sim 2$.
The evaluated penetration length is longer than the value ($\ell = 1/\alpha \sim 120$~nm) evaluated from the absorption coefficient of the BI phase, as shown in Fig.~\ref{fig:R2}(c).
This indicates that PE creates a new PI metastable state with a different absorption coefficient from BI. 
This suggests that the electronic state of the PI phase is not identical to that of BI, which is consistent with the different refractive index, as described before.
This expansion of the volume of the PI state inside into the sample after PE is namely as ``domino effect'' and is known as one of the unique non-linear phenomena of PI phase transitions~\cite{Koshino1998-pd}.
Additionally, as shown in Fig.~\ref{fig:R2}(b), $\tau_v$ decreases to about half as the laser power increases.
This fact suggests that an absorption coefficient slightly increases, representing a slight change in the electronic state of the PI phase from that of BI, {\it i.e.}, the density of states near the Fermi level increases.
However, the absorption coefficient has no significant change after PE, suggesting no significant metallic phase transition.
This implies that the electronic structure change is insignificant, consistent with the narrowing of the energy gap from $\sim 100$~meV to $\sim50$~meV~\cite{Chen2022-fn}.

Figure~\ref{fig:R2}(d) shows a schematic figure of the creation and relaxation dynamics of the PI state after PE.
Under a weak PE, electrons at each site are excited individually and relax immediately ($\leq 1$~ps), which is the origin of the fast component.
If the nearby sites are excited simultaneously, the PI states could form domain structures.
The relaxation time of such domains becomes as long as 10~ps, which is the origin of the slow component, $\tau_s$.
Under strong PE, many sites are simultaneously excited to form large domains.
These states are PI metastables with a long relaxation time ($>190$~ps) described as the residual component.
The electronic structure of the PI metastable state is different from that of the BI state, as discussed before.
To justify the above speculation, the following section investigates the electronic configuration of Sm ions in the PI state using the Sm~$3d$~XAS after PE.

%%%%%%%%%%%%%%%%%%%%%%%%%%%%%%%%%%%%%%%%%%%%%%%%%%
%%%%%%%%%%%%%%%%%%%%%%%%%%%%%%%%%%%%%%%%%%%%%%%%%
\subsection{Pump-probe X-ray absorption at Sm~$3d$ edge}

%%%%%%%%%%%%%% FIG. 3. trXAS %%%%%%%%%%%%%%%%%%%%
\begin{figure}[t]
\begin{center}
\includegraphics[width=0.45\textwidth]{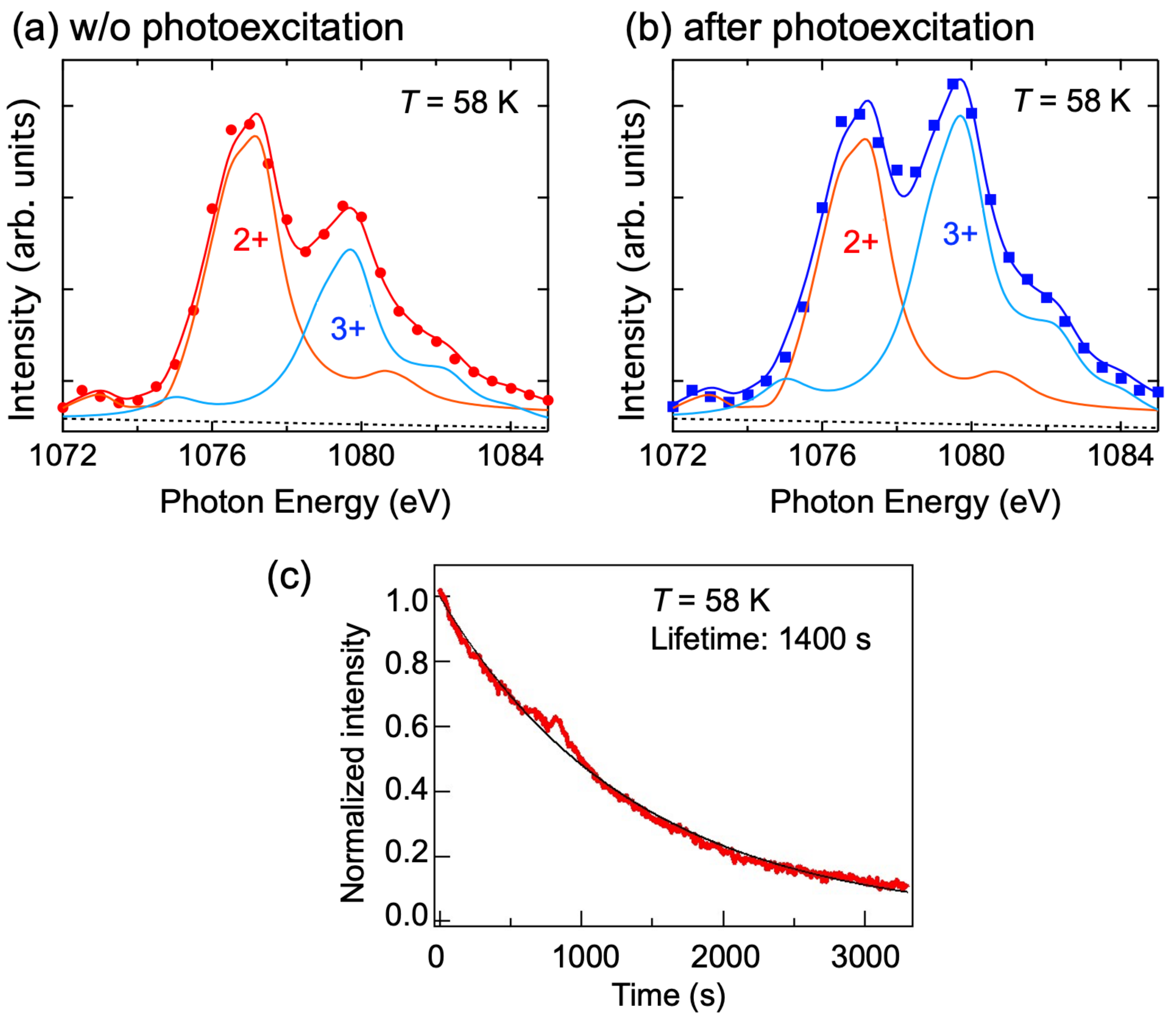}
\end{center}
\caption{
(a, b) X-ray absorption spectra (marked lines) at the Sm~$3d_{5/2}$ core level without (a) and after (b) photoexcitation for the time of 300~s taken at the temperature of $58$~K.
Solid lines are the \Smdi and \Smtri components derived from the calculated absorption spectra, and dotted lines are backgrounds from other components.
(c) Temporal structure of the 1080-eV peak after stopping of PE.
The solid curve is an exponential decay curve with a lifetime of 1400~s.
The vertical axis is normalized between (a) and (b).
}
\label{fig:XAS}
\end{figure}
%%%%%%%%%%%%%%%%%%%%%%%%%%%%%%%%%%%%%%%

Laser-pump--X-ray-absorprion-probe spectroscopy was performed to clarify the change in electronic structure and the valence in the residual components (PI phase) speculated with the time-resolved reflectivity measurement.
Figure~\ref{fig:XAS}(a) shows the Sm~$3d_{5/2}$ XAS spectrum of BI, Sm$^{2+}$ and Sm$^{3+}$ peaks appear at 1077 and 1080~eV, respectively, at the temperature of 58~K.
After one-laser-pulse PE, we could not observe the spectrum change within the experimental accuracy of about 1~\%, suggesting a very small valence change by one pulse owing to the different penetration depth between the laser pulse ($\sim 100$~nm) and the X-ray ($\sim 1000$~nm) and the time width of the X-ray.
According to our time-resolved reflectivity, though a 20-\% change in \dR was observed just after the PE, it relaxed to less than 5~\% within 50~ps of the X-ray pulse width.
Therefore, the invisible spectral change in the XAS resulting from one laser pulse is roughly consistent with a very small change in the optical constants after PE observed in our time-resolved reflectivity described before.

From the time-resolved reflectivity, it is known that the residual component remains more than 190~ps but disappears at 2~ms at 300~K.
However, the relaxation time increases as the material is cooled to lower temperatures.
If the next laser pulse comes to a sample before a complete relaxation, the volume fraction of the PI state increases, which is called a pile-up effect. 
At the temperature of $58$~K, we could significantly change the XAS spectra by a pile-up effect.
As shown in Fig.~\ref{fig:XAS}(b), the spectrum changes after the laser pulse is applied for 300~s ($3\cdot 10^5$~pulse) at 58~K.

Before PE shown in Fig.~\ref{fig:XAS}(a), not only the peak at 1077~eV (mainly originating from a \Smdi component) but also the peak at 1080~eV (mainly from \Smtri) was observed, but the \Smdi peak is dominant.
The mean valence before PE can be evaluated as $2.32 \pm 0.01$ from the peak intensity ratio at 1077~eV and 1080~eV in Fig.~\ref{fig:XAS}(a) deconvoluted with calculated Sm~$3d$ absorption spectra~\cite{Kaindl1984-kw}.
The obtained mean valence is consistent with that evaluated with the \Smtri~$3d$~XPS spectra~\cite{Nakamura2023-np}.
Just after PE shown by blue circles in Fig.~\ref{fig:XAS}(b), the intensity of the \Smtri peak increased compared to that of \Smdi, the mean valence increased to $2.46 \pm 0.01$ after PE.
Figure~\ref{fig:XAS}(c) shows the intensity decay of the 1080-eV peak, indicating the recovery to the original valence number before PE.
It should be noted that the recovery to nearly \Smdi suggests the intrinsic valence change with PE because the other effects, such as the sample damage by oxidization and the desorption of S atoms, are considered to induce the change to \Smtri, the opposite behavior.
A long lifetime ($\sim 1400$~s) is consistent with the pile-up effect of PE.
The result indicates that the PI valence transition appears in SmS after many laser-pulse irradiations, which may suggest the possibility of the origin of excitonic instability for the BGT.
When the BGT occurs, the lattice contraction should appear.
It was also investigated after PE and described in the next section.
%However, the mean valence of the PI phase does not reach 2.7 of the GM phase~\cite{Imura2020-zq}.
%The reason is discussed in the discussion section.

%%%%%%%%%%%%%%%%%%%%%%%%%%%%%%%%%%%%%%%%%%%%%%%%%%
%%%%%%%%%%%%%%%%%%%%%%%%%%%%%%%%%%%%%%%%%%%%%%%%%
\subsection{Pump-probe X-ray diffraction}

%%%%%%%%%%%%%% FIG. 4. trXRD %%%%%%%%%%%%%%%%%%%%
\begin{figure}[t]
\begin{center}
\includegraphics[width=0.48\textwidth]{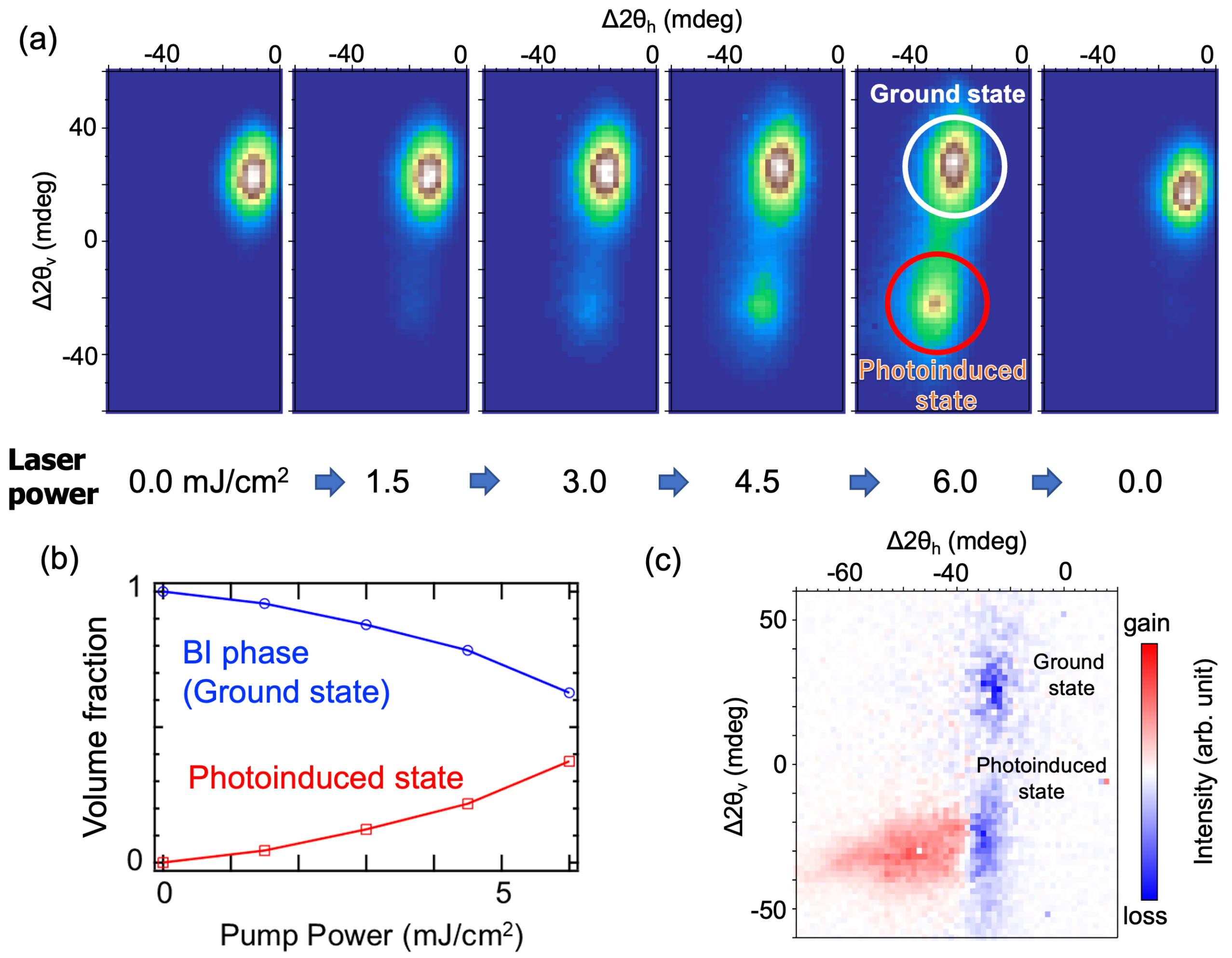}
\end{center}
\caption{
Excitation-laser-power dependence of X-ray diffraction of the (400) Bragg peak after PE in the time of $0 < t < 0.1~{\rm ns}$.
(a) (400)-Bragg peak image as a function of the excitation laser power from 0.0 to 6.0~\mJ.
The final image was taken without laser irradiation after the PE experiment.
(b) The volume fraction (Bragg peak intensity) of the PI phase and the BI phase (Ground state) as a function of excitation laser power.
(c) Subtraction image of the (400) Bragg peak just before PE ($t \sim -0.1$~ns) from that after PE ($t \sim 0.1$~ns).
}
\label{fig:XRD}
\end{figure}
%%%%%%%%%%%%%%%%%%%%%%%%%%%%%%%%%%%%%%%

Figure~\ref{fig:XRD} shows the excitation intensity dependence of the (400) Bragg peak just after PE.
The horizontal axis $\Delta2\theta_h$ (the vertical axis $\Delta2\theta_v$) represents the relative angle in the direction of (perpendicular to) the diffraction angle.
The minus angle direction of $\Delta2\theta_h$ indicates a lower diffraction angle (longer lattice constant).
In the condition without pump laser pulse (0.0~\mJ), a single Bragg peak was visible, as shown in Fig.~\ref{fig:XRD}(a).
With increasing excitation laser power, a new peak indicated by the red circle appeared at a different $\Delta2\theta_v$ position from the original peak, and the intensity gradually increased.
After the excitation laser was turned off, the newly formed peak disappeared and returned to its original image.
These facts imply that the newly appeared peak is the (400) Bragg peak of the PI phase.

The diffraction angle $\Delta2\theta_h$ of the PI phase is smaller by 5~mdeg than that of the BI phase, which corresponds to the lattice constant increases by $7 \cdot 10^{-4}$~\Ang ($\sim 0.01$~\% of the lattice constant of SmS).
This result suggests that the lattice constant of the PI phase is almost the same as that of the BI phase.
Besides the $\Delta2\theta_h$ direction, the two Bragg peaks are located at different positions in the $\Delta2\theta_v$ direction perpendicular to the diffraction angle.
Since lattice expansion and contraction cannot explain this effect, its origin is discussed later.

To further investigate the excitation power dependence of the Bragg peak, we plotted the volume fractions (integration of the peak intensities) of the PI phase and the BI phase (ground state) as a function of excitation laser power in Fig.~\ref{fig:XRD}(b) obtained from Fig.~\ref{fig:XRD}(a).
The volume fraction of the PI phase (the BI ground state) increases (decreases) nonlinearly with increasing laser power and the sum of them is almost constant, suggesting that a part of the ground state changes to the PI phase.
At the highest excitation power of 6.0~\mJ, $40$-\% area changes to the PI phase.
Such a nonlinear change usually appears as a general behavior of PI phase transitions of other PI phase-transition materials~\cite{Nasu2004-px}.
It should be noted that the laser power to the sample damage in the reflectivity measurement in this work is above 4~\mJ, but in the XRD measurement, the sample was not damaged by the laser power of 6~\mJ.
This inconsistency may be due to the different experimental setups, such as the different incident angles and focusing conditions.
Therefore, the laser power in the reflectivity and the XRD may be mismatched.

To investigate the time evolution of the Bragg peak in detail, we took the subtraction image (Fig.~\ref{fig:XRD}(c)) of that before PE ($t \sim -0.1$~ns) from that after PE ($t \sim 0.1$~ns).
Just after PE, the diffraction intensity of both the PI phase and the BI ground state decreases, and the low-angle side of the peak of the PI phase increases.
The origin of the red area's appearance is considered a thermal expansion effect.
The BI ground state has no thermal effect because all photo-excited areas changed to the PI state.
Due to the different penetration depths, the remaining BI component is not irradiated with an excitation laser pulse and is not thermally expanded.
Therefore, there is no red area, but the intensity of the original peak decreases.
The peak intensity moved to the red area in the PI state.
This suggests that all photo-excited areas in BI moved to the PI state, and the lattice of the PI phase near the surface, which is directly exposed to the laser pulse, thermally expands instantaneously ($t < 0.1$~ns) by PE.

The positions of the Bragg peaks of the PI and BI states shifted to the low-angle side depending on the excitation intensity, suggesting a thermal expansion over the entire sample under laser irradiation.
At the highest excitation intensity of 6.0~\mJ shown in Fig.~\ref{fig:XRD}(a), the lattice constant of the whole sample is increased by about $2.1 \cdot 10^{-3}$~\Ang ($\Delta2\theta_h \sim 15$~mdeg) due to the thermal effect of laser irradiation.
In comparison with the thermal expansion of the lattice constant by 0.12~\% (0.007~\Ang) between 100 and 300~K~\cite{Matsubayashi2007-bq}, the average temperature increase across the entire sample due to laser irradiation can be evaluated as $\sim 30$~K at the highest laser power of 6.0~\mJ.
The temperature increase of 30~K is not as high as the temperature for thermally induced phase transitions~\cite{Pohl1974-qf,Travaglini1984-pq}.

The Bragg peaks of the PI and BI states are observed at different $\Delta2\theta_v$, and this change corresponds to an angular change of about 40~mdeg.
Since the vertical angle change cannot be explained with the lattice constant change, one possibility for the different $\Delta2\theta_v$ is the sample surface of the PI phase was tilted by about 20~mdeg from the BI ground state.
The surface tilting randomly appears in a crystal due to the PI phase transition and thermal expansion.
Since such an effect does not occur only in a specific direction but appears in any direction, it is possible that the same effect occurs in the direction of the diffraction angle.
Assuming an increasing angle by 20~mdeg in maximum along the diffraction angle direction, we can estimate that the change of the lattice constant of the PI state from BI is less than $3 \cdot 10^{-3}$~\Ang (0.05~\% of the lattice constant of BI) in maximum.
This change is much smaller than the lattice constant change of about 5~\% by the BGT~\cite{Imura2020-zq}.
Therefore, almost no lattice contraction occurs in the PI phase transition of SmS.
This suggests that the PI phase is inconsistent with GM but similar to BI.
The result is consistent with the reflectivity and XAS measurements.

%%%%%%%%%%%%%%%%%%%%%%%%%%%%%%%%%%%%%%%
\section{Discussion}

The reflectivity, XAS, and XRD experiments after PE provide information on the changes in optical constants, valence, and lattice constants due to PE, respectively.
In the reflectivity measurements, fast, slow, and residual relaxations are observed with PE, with the residual relaxation having a time constant of more than a hundred ps.
A vibration component was also observed in the slow relaxation time scale, indicating that the PI phase expands into the crystal by a sound velocity.
However, the Drude component observed in GM does not appear in the PI phase, suggesting that the PI phase is a newly appeared metastable state different from the GM phase.
The optical constants, the refractive index and absorption coefficient, of the PI state are different from those of the BI state, implying that the electronic structure of the PI phase changes from that of BI.

Furthermore, with increasing laser power, the relaxation time of the vibrational component ($\tau_v$) of the reflectivity decreases, suggesting an increase in the absorption coefficient.
This fact suggests that the electronic density of states near the Fermi level increases with the application of laser power. which is consistent with the narrowing of the energy gap observed by a time-resolved angle-resolved photoemission spectroscopy~\cite{Chen2022-fn}.

In the Sm~$3d$~XAS with PE, the average valence of Sm increases from $2.32 \pm 0.01$ before PE to $2.46 \pm 0.01$ after PE.
This means the electronic state approaches GM after PE but does not reach 2.7 of the GM phase and saturates~\cite{Imura2020-zq}.
Taken together with the change in reflectivity, this suggests that the electronic state moves to a novel PI phase, which differs from the BI ground state and the pressure-induced GM phase.

Although the lattice constant in the GM phase is known to be about 5~\% smaller than that in the BI phase~\cite{Imura2020-zq}, XRD measurements after PE show that the lattice constant of the PI phase changes only less than 0.1~\% from that of the BI phase.
This fact also means that the PI phase differs entirely from GM.

%%%%%%%%%%%%%% FIG. 5. Schematic figure %%%%%%%%%%%%%%%%%%%%
\begin{figure}[t]
\begin{center}
\includegraphics[width=0.48\textwidth]{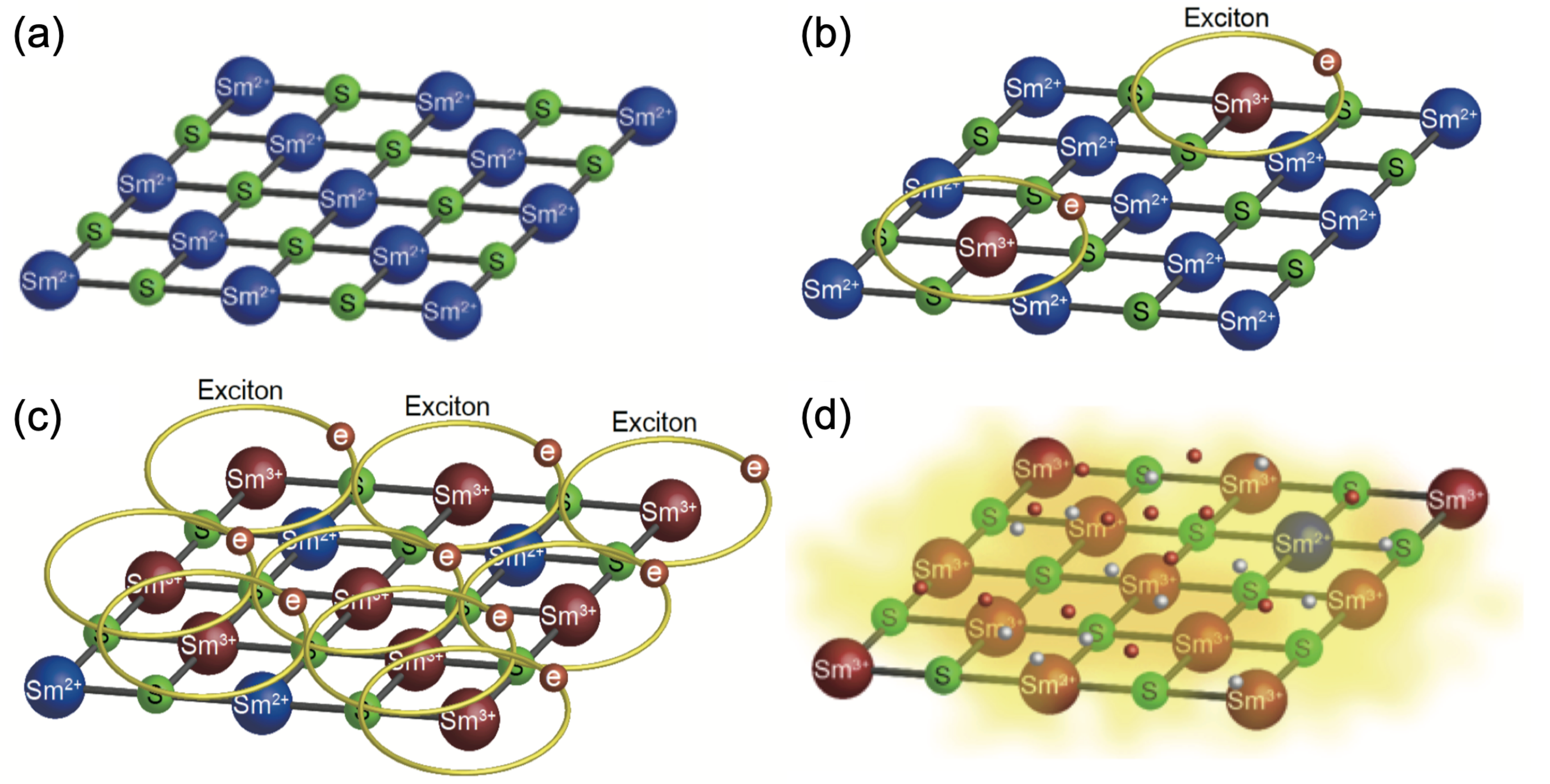}
\end{center}
\caption{
Schematic figure of the PI phase transition along the excitonic instability picture.
(a) The BI phase.
(b) After the PE by several pulses irradiated to the sample, $4f$ electrons are excited and become excitons. 
The created excitons are isolated and localized at the original sites.
The electronic structure and optical constants are slightly changed, but the lattice constant is identical before the PE.
These changes were observed by reflectivity and XRD, respectively.
(c) After many-laser-pulse PE with a pile-up effect, many excitons are created but still isolated.
The valence transition becomes visible by XAS because of many \Smtri states.
(d) The GI phase, where excitons are condensed, and the state becomes metallic.
}
\label{fig:manga}
\end{figure}
%%%%%%%%%%%%%%%%%%%%%%%%%%%%%%%%%%%%%%%

These results indicate that BGT cannot be induced simply by generating excitons due to PE.
Schematic figures for explaining the experimental results in this paper are shown in Fig.~\ref{fig:manga}.
The reflectivity and XRD show that after the PE by several laser pulses to BI (Fig.~\ref{fig:manga}(a)), Sm~$4f$ electrons are excited and become excitons localized at the original sites and isolated, as shown in Fig.~\ref{fig:manga}(b).
Then, the electronic structure and optical constants slightly change, but the lattice constant is identical before PE.
After being excited by many laser pulses at a low temperature, many PI states are generated by the pile-up effect, as illustrated in Fig.~\ref{fig:manga}(c).
Then, many excitons are created, and a valence change can be observed.
In the excitonic instability picture, the state can be regarded as a BEC state, in which electron-hole pairs are created but isolated~\cite{Varma1976-op,Shao2024-vm}.
However, the phase cannot reach GM, where excitons are condensed as a BCS state and become free carriers, as shown in Fig.~\ref{fig:manga}(d).
This suggests that the BEC-BCS transition of SmS cannot be realized only by PE.

Other studies have attempted to induce BGT by manipulating the electronic state.
One is the observation of nonlinear conduction and its electronic state when an external current is introduced into BI~\cite{Ando2020-fu,Kimura2024-jz}.
The current-induced insulator-to-metal transition appears, and the mean valence approaches the trivalent state where the current flow enhances the $d$-$f$ hybridization.
However, it was found that even if the current was applied until the sample was broken, the sample color did not change, and a significant Drude component did not appear, suggesting the absence of BGT.
Furthermore, it has been reported that when electrons are injected by an alkali metal deposition in an ultrahigh vacuum, the sample surface state becomes trivalent, but metallization does not occur~\cite{Nakamura2023-np}.
Common to all results obtained with the different perturbation methods described above is that a change in electronic state is observed, but the lattice constant does not change.
The fact that the ionic radius of \Smtri is smaller than that of \Smdi is the origin of the lattice contraction from BI to GM.
However, the lattice contraction needs the time scale of $10^{-9}-10^{-10}$~s that is much longer than that of the electronic excitation time scale of $\sim 10^{-15}$~s.
The PE by the laser pulse used here is also short ($\sim 10^{-13}$~s), so the lattice contraction cannot occur during the laser irradiation.
Therefore, this suggests that BGT is not caused by an electronic state change alone but requires lattice contraction.

%%%%%%%%%%%%%%%%%%%%%%%%%%%%%%%%%%%%%%%
\section{Conclusion}
To investigate the origin of the pressure-induced insulator-to-metal transition (BGT) in SmS and the relation to an expected excitonic instability, the reflectivity, X-ray absorption, and X-ray diffraction after photoexcitation (PE) have been measured to obtain information on the change in optical constants, lattice constants, and the Sm valence, respectively.
Reflectivity measurements showed a 22-\% decrease at \hn $\sim 1.55$~eV, a very long relaxation time ($> {\rm several}~100$~ps) after PE, and the propagation of the photo-induced (PI) phase within the crystal.
The Sm valence was also observed to increase from 2.32 to 2.46 when laser pulses were applied.
These results suggest that the electronic state of the PI phase is different from that of the BI ground state, which is consistent with the gap narrowing observed in the time-resolved photoelectron spectroscopy previously reported.
However, the lattice constants in the PI phase are almost consistent with those of the BI state within less than 0.1~\%, which is entirely different from the GM phase.
These results imply that the PI phase after PE differs entirely from the GM phase.
Even though the mean valence slightly increases with PE, the PI state is inconsistent with GM.
This suggests that the BGT cannot be achieved solely by creating excitons by PE; other effects, such as lattice contraction, are needed. 

%%%%%%%%%%%%%%%%%%%%%%%%%%%%%%
\section*{Acknowledgments}
We appreciate the technical support of the staff members of SPring-8.
This work was performed at SPring-8 with the approval of the Japan Synchrotron Radiation Research Institute Proposals Nos.~2022A1097, 2019B7455, 2019A7453 and was partly supported by JSPS KAKENHI (Grant Nos.~20H04453, 23H00090, 23K04568, 24K21197).

%%%%%%%%%%%%%%%%%%%%%%%%%%%%%%
%\begin{thebibliography}{99}
%
%
%\end{thebibliography}

%\bibliographystyle{plain}
\bibliographystyle{apsrev4-1}
\bibliography{SmS_PI_Notes}

\end{document}